\documentclass[12pt]{article}
\usepackage{amsmath, amsthm, amsfonts, amssymb, amsxtra}
\usepackage{graphicx,psfrag,epsf}
\usepackage{enumerate}
\usepackage{natbib}
\usepackage{url} 
\usepackage{bm, bbm}
\usepackage{booktabs}
\usepackage[pagebackref=false]{hyperref} 
\usepackage[svgnames]{xcolor}
\usepackage{subcaption}
\usepackage{physics} 
\usepackage[margin = 1in]{geometry}
\usepackage{setspace}
\usepackage{multirow} 
\usepackage{xr} 
\usepackage{endfloat}

\allowdisplaybreaks[1]

\usepackage{color}
\definecolor{blue2}{rgb}{0.1, 0.2, 0.6}

\theoremstyle{definition}

\newcommand{\R}{\rm I\!R}
\newcommand{\E}{\mathbbm{E}}

\hypersetup{
    colorlinks = true,
    linkcolor  = blue,
    urlcolor   = blue,
    citecolor  = blue,
    pdftitle   = {A unified framework for spatial data} 
  }
  
\author{Lucas da Cunha Godoy\thanks{Department of Statistics, University of
      Connecticut}\thanks{Department of Ecology and Evolutionary Biology,
      University of California Santa Cruz}\\ \texttt{ldcgodoy@ucsc.edu} \and
    Marcos Oliveira Prates\thanks{Department of Statistics, Universidade Federal
      de Minas Gerais} \and
    Jun Yan\footnotemark[1]}

\begin{document}

\title{Statistical Inferences and Predictions for Areal Data and Spatial Data
  Fusion with Hausdorff--Gaussian Processes}

\maketitle

\begin{abstract}
  Accurate modeling of spatial dependence is pivotal for both parameter
  estimation and prediction in spatial data analysis. The spatial structure of
  the data significantly narrows the set of available models. Existing areal
  data models often ignore polygon geometry, while data fusion models require
  computationally intensive integration, limiting their application. In response
  to these issues, we propose the Hausdorff-Gaussian process~(HGP), a versatile
  model utilizing the Hausdorff distance to capture spatial dependence in both
  point and areal data. Integration into generalized linear mixed-effects models
  enhances its applicability, particularly in addressing data fusion
  challenges. We validate our approach through a comprehensive simulation study
  and application to two real-world scenarios: one involving areal data and
  another demonstrating its effectiveness in data fusion. Results show the HGP
  is a competitive, flexible, and robust solution for modeling diverse spatial
  data, with potential applications in fields like public health and climate
  science.

\end{abstract}


\section{Introduction}\label{sec:intro}

Accounting for spatial dependence is pivotal in spatial data analysis. The
nature of the observed spatial data significantly influences the choice of
statistical methodology. Point-referenced data are typically analyzed using
geostatistics methods~\citep[e.g.,][]{cressie1993statistics}. For areal data,
the conditional autoregressive model~\citep[CAR]{besag1974spatial} and its
variants~\citep{besag1991bayesian, rodrigues2012bayesian, datta2019spatial,
  cruz2023inducing} are often preferred. These differences become increasingly
critical in complex inferential problems, such as predicting a variable in a
target map using data from a source map. This issue, known as change of
support~\citep[pg.~284--289]{cressie1993statistics}, requires tailored methods
depending on the spatial units involved~\citep{gelfand2001change}. Spatial data
from diverse sources often differ in spatial units or supports, resulting in
spatial misalignment~\citep{gotway2002combining}. Data fusion is necessitated
when a spatial variable is provided by multiple, potentially misaligned,
sources~\citep{moraga2017geostatistical}.

Both traditional areal models and data fusion models for spatial data exhibit
inherent limitations. Traditional areal models lack interpretable spatial
dependence parameters~\citep{assuncao2009neighborhood}. An exception of that
last limitation is the directed acyclic graph auto-regressive~(DAGAR) model,
which offers an interpretable parameter~\citep{datta2019spatial}. These models,
which rely on a fixed adjacency graph, are not suited for out-of-sample
prediction, as introducing new spatial units fundamentally alters the whole
model structure. Furthermore, when overlapping spatial units are
present~\citep{bakar2020areal}, another limitation is that they do not account
for the size or shape of the areal units. Data fusion models, which view areal
data as averages of a continuous process, may be unsuitable for discrete
outcomes~\citep{gelfand2001change}. Additionally, their reliance on arbitrary
discretization for likelihood evaluation lacks consensus on optimal
resolution~\citep{fuentes2005model, liu2011empirical, wang2021combining} and can
introduce unquantifiable biases~\citep{goncalves2018exact}.

We introduce a unified approach for areal, point-referenced, and mixed spatial
data by combining Gaussian processes~(GPs) with the Hausdorff
distance~\citep{arutyunov2016some}, which seamlessly handles these diverse
formats in a single framework. The Hausdorff distance, defined on a reference
metric space, effectively measures the distance between two sets while
accounting for their shape. For point pairs, it reduces to the distance metric
associated to the reference metric space. GPs, recognized for their flexibility,
are widely utilized in spatial data modeling, notably as priors in generalized
mixed effects models~\citep{diggle1998model}, in spatially varying regression
coefficients~\citep{gelfand2003spatial}, generalized multilevel
models~\citep{prates2024fast}, and in addressing spatial misalignment
issues~\citep{banerjee2002prediction}. We propose the Hausdorff-Gaussian
process~(HGP), a GP with a correlation function defined on the Hausdorff
distance between spatial units. This allows us to directly incorporate shape and
size information into the analysis. A crucial aspect of our work is to assess
the competitiveness of the HGP, particularly when compared to specialized models
designed for areal and mixed spatial data. We investigate whether the HGP, using
a powered exponential correlation, can effectively model areal and mixed spatial
data, thereby offering a novel and versatile tool for analyzing spatial data
with mixed supports.

This paper is structured as follows. Section~\ref{sec:hgp} introduces the
Hausdorff distance, discussing its properties, and then delineates the HGP,
highlighting its significant applications. Then, in Section~\ref{sec:bayes}, we
set up spatial generalized linear mixed models (GLMM) with HGP components and
detail their inferences in the Bayesian framework. A simulation study to compare
the proposed model with specific models for areal and fused data is presented in
Section~\ref{sec:sim}. Section~\ref{sec:data} demonstrates the practical utility
of our model through two distinct applications, one on areal data and the other
on data fusion. Section~\ref{sec:disc} concludes with a discussion. For
completeness, we introduce models designed specifically for areal and fused data
in the Appendix. The Supplementary Material and the code to reproduce the
analyses are available in the following GitHub repository:
\url{https://github.com/lcgodoy/jabe-d-25-00205}.

\section{Hausdorff--Gaussian Process}\label{sec:hgp}

\subsection{Hausdorff Distance}\label{sec:hausdist}

The Hausdorff distance is a metric measuring the distance between two
sets. Specifically, it is the greatest distance from a point in one set to the
closest point in the other set. Consider a metric space~$(D, d)$, where~$D$ is a
spatial region of interest and~$d(x, y)$ defines a distance between any two
elements~$x, y \in D$. For a single point~$x$ and a set~$A$,
define~$d(x, A) = \inf_{a \in A} d(x, a)$ as the closest distance from~$x$ to any
element in~$A$. The directed Hausdorff distance from a set~$A$ to a set~$B$
is~$\vec{h}(A, B) = \sup_{a \in A} d(a, B)$.  This measure is not necessarily
symmetric. The symmetric Hausdorff distance is defined as the greater of the two
directed Hausdorff distances:
\[
  h(A, B) = \max \left \{ \vec{h}(A, B), \vec{h}(B, A) \right \}.
\]
Note that if~$A$ and~$B$ are both singletons, then~$h(A, B) = d(A, B)$. If they
are compact, there exist $a \in A$ and $b \in B$ such
that~$h(A, B) = d(a, b)$~\citep{arutyunov2016some}.

In spatial statistics, sample units~(e.g., points, polygons) are repsented as
non-empty, closed, and bounded subsets of a study region
$D \subset \mathbb{R}^k$. In Euclidean domains, the Heine-Borel theorem asserts that closed and
bounded sets are compact~\citep[pg.~98]{abbott2015understanding}. The Hausdorff
distance is a suitable metric for the collection of these non-empty compact
subsets, hereafter denoted
$\mathcal{B}(D)$~\citep[Appendix~C]{molchanov2005theory}. By the properties of a metric,
$h(A, B) = 0$ if and only if $A$ and $B$ are identical. The metric space induced
by the tuple~$(\mathcal{B}(D), h)$ inherits properties like compactness and completeness
from~$(D, d)$.  Further details regarding the topology induced by the Hausdorff
distance can be found in Appendix C in \citet{molchanov2005theory}. Lastly, in
the Supplementary Material, Section~S.1, we provide a toy example comparing the
Hausdorff distance to the distance between centroids and, in addition, between
borders.

\subsection{Hausdorff--Gaussian Process}

HGP is a natural extension of GP for modeling spatial data defined on the
Hausdorff distance. A GP is a stochastic process over an index set such that all
its finite-dimensional marginal distributions are multivariate Normal
distributions. Let~$D$ be a spatial region of interest. An HGP index set is
taken to be the class of non-empty compact subsets of~$D$, denoted
$\mathcal{B}(D)$. This class encompasses both singletons and polygons, which can be
represented as closed and bounded sets. To define an isotropic HGP~$Z(\cdot)$
over~$\mathcal{B}(D)$, we specify a marginal mean function~$m(\cdot)$, a positive-valued
marginal standard deviation~(SD) function~$v(\cdot)$, and a valid isotropic
correlation function~$r(h)$ for any two sets apart in Hausdorff
distance~$h$. With these functions, an HGP is a stochastic process~$Z(\cdot)$ whose
marginal distribution
at~$\{\mathbf{s}_1, \ldots, \mathbf{s}_{n}\} \in \mathcal{B}(D)$ is multivariate Normal with mean
vector and covariance matrix, respectively,
\begin{equation}
  \label{eq:hgp}
  \E[Z(\mathbf{s}_i)] = m(\mathbf{s}_i) \quad \text{and} \quad
  \mathrm{Cov}[Z(\mathbf{s}_i), Z(\mathbf{s}_j)] =
  v(\mathbf{s}_i) v(\mathbf{s}_j) r\big(h(\mathbf{s}_i, \mathbf{s}_j)\big),
\end{equation}
for~$i, j = 1, \ldots, n$, where~$h(\mathbf{s}_i, \mathbf{s}_j)$ is the Hausdorff
distance between~$\mathbf{s}_i$ and~$\mathbf{s}_j$. This process allows for
representations of diverse spatial data, from areal to mixed, making it suitable
for a wide range of applications such as disease mapping and data fusion.

Denote the HGP process defined in Equation~\eqref{eq:hgp} by
$\textrm{HGP}\{m(\cdot), v(\cdot), r(\cdot)\}$. A GP's mean function has little impact on
the process' local behavior~\citep[pg.~12]{stein1999interpolation}. They capture
large-scale variations, and, hereafter~$m(\cdot)$ is either held constant or defined
through covariates. The SD function~$v(\cdot)$ accommodates both homoscedastic and
hereroscedastic scenarios. For instance, with a log link, a natural model is
\begin{equation}
  \label{eq:nu}
  \log v(\mathbf{s}) = \alpha_0 + \alpha_1 w(\mathbf{s}),
\end{equation}
where~$w(\mathbf{s})$ is a covariate available for any~$s \in
\mathcal{B}(D)$. Equation~\eqref{eq:nu} allows a distinction between point-referenced and
areal data, a need in data fusion
applications~\citep{wang2021combining}. Specifically, to enable different
variances between these data types, we can
define~$w(\mathbf{s}) = \mathbbm{1}(\mathcal{A}(\mathbf{s}) > 0)$,
where~$\mathcal{A}(\cdot)$ is a function that returns the area of a spatial object. Certainly,
care is necessary in specifying~$v(\mathbf{s})$ to ensure a valid
GP~\citep{palacios2006non}.

\subsection{Correlation Function}\label{sec:corr}

The correlation function~$r(\cdot)$ needs to be carefully chosen to ensure the
validity of the HGP. In particular, we focus on isotropic correlation
functions. Thus, unless explicitly mentioned, all subsequent references to
correlation functions will assume they are isotropic. A valid correlation
function must be positive definite and meet two key criteria:~$r(0) = 1$
and~$\lim_{h \to \infty}r(h) = 0$~\citep[Ch.~2.3.2]{cressie1993statistics},
where~$h$ is the Hausdorff distance between a
pair~$\mathbf{s}, \mathbf{s}' \in \mathcal{B}(D)$. The latter ensures that spatial
correlation diminishes with increasing distance. It is noteworthy that while
most geostatistical correlation functions are constructed for Euclidean
spaces~\citep[Ch.~2.5.1]{cressie1993statistics}, their positive definiteness may
not extend to non-Euclidean spaces~\citep{curriero2006use}.

For practical purposes, we propose to use a two-parameter powered exponential
correlation~(PEC) function
\[
  r(h; \varphi, \nu) = \exp \left \{ - {\left(\frac{h}{\varphi}\right)}^{\nu}\right \},
\]
where~$\nu$ controls the smoothness and~$\varphi$ governs the range of the spatial
dependence. To enhance interpretability, we reparametrize this function
with~$\rho = {\log(10)}^{1 / \nu} \varphi$. Consequently, $\rho$ is the distance at which the
spatial correlation between a pair of spatial units drops to~$0.10$, commonly
referred to as the practical range. The PEC function is appealing because,
for~$\nu \in (0, 1]$, it is positive definite as long as applied to a conditionally
negative definite~(CND) distance function~\citep{schoenberg1938metric}. However,
to the best of our knowledge, no theoretical results establish when the
Hausdorff distance is CND, nor are there results concerning positive definite
functions of this distance. Thus, we recommend empirically assessing the
positive definiteness of the PEC function before analyzing the data.

To this end, we empirically assess the positive definiteness of the PEC function
for~$\nu \in (0, 1]$. We first calculate Hausdorff distance matrices for the areal
and fused data to be analyzed in Sections~\ref{sec:data_areal}
and~\ref{sec:data_fusion}, respectively. For comparison purposes, we re-scale
these distance matrices so that all their elements are less than or equal
to~1. We then evaluate the PEC function for these matrices using a grid of
values for~$\nu \in (0, 1]$ and~$\rho \in (0, 1.5]$, where the upper bound
of~$\rho$ represents~1.5 times the largest distance observed in the respective
matrix. This process generates a set of corresponding correlation
matrices. Finally, we compute the smallest eigenvalue for each of these
correlation matrices. A negative eigenvalue indicates the corresponding PEC
function is not positive definite for that combination of parameters. However,
even valid covariance functions can be ill-conditioned or nearly rank-deficient
at dense locations or when there exists a strong spatial
dependence~\citep{banerjee2013efficient}.

\begin{figure}[t]
  \centering
  \includegraphics{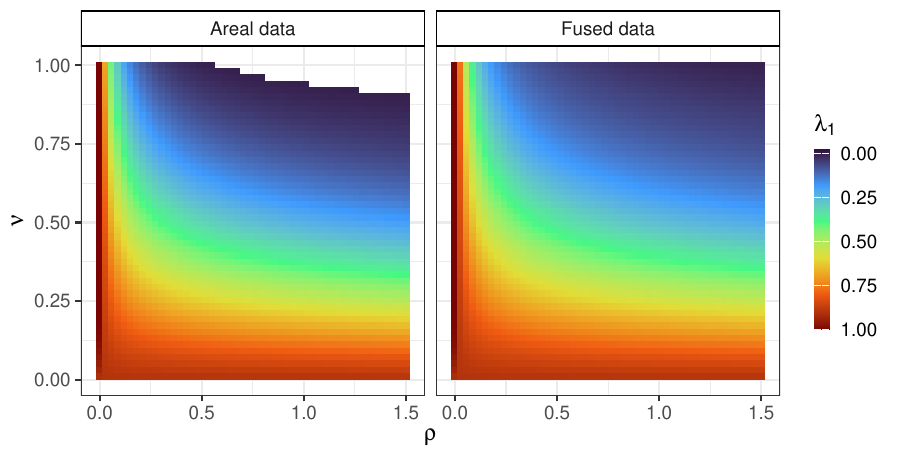}
  \caption{Smallest eigenvalue of the powered exponential correlation matrix
    under different combinations of smoothness~($\nu$) and practical
    range~($\rho$) for areal and fused data. The transparency indicates regions
    where $\lambda_1 < 0$.}\label{fig:corr-exp}
\end{figure}

Figure~\ref{fig:corr-exp} displays the smallest eigenvalues~($\lambda_1$) of the
PEC-induced correlation matrices for the fused and areal datasets from
Sections~\ref{sec:data_areal} and~\ref{sec:data_fusion}, respectively. The fused
data exhibits positive eigenvalues for all~$(\nu, \rho)$ pairs. In contrast, the
areal data reveals some small negative eigenvalues~(range $-0.0099$ to
$-0.00006$) under strong spatial dependence and at~$\nu \approx 1$, the border of the
parametric space. Based on these empirical results, it is safe to use the PEC
function with~$\nu \in (0, 1]$ for fusion data, but, for the areal application, it
may be advisable to stick with a~$\nu$ value less than~1. We further investigate
the PD property of the PEC function across different datasets and
domains. Similiar conclusion are obtained and these results are presented in
Section~S.2 of the Supplementary Material.

\section{HGP in Spatial GLMM}\label{sec:bayes}

\subsection{Model}\label{sec:spglm}

Generalized linear models for spatial data can benefit from replacing the GP
with the HGP as spatial random effects. In this context, assume
$\mathcal{S} = \{ \mathbf{s}_1, \ldots, \mathbf{s}_n \}$ is a set of spatial units such that
$\mathbf{s}_i \in \mathcal{B}(D)$, where $D \subset \R^2$. Consider a response variable
$Y(\mathbf{s}_i)$ and a $p$-dimensional vector of covariates $\mathbf{X}_i$
associated to the $i$-th spatial unit. Let
$\mathbf{Z} = {\big[Z(\mathbf{s}_1), \ldots, Z(\mathbf{s}_n)\big]}^\top$ be an
$n$-dimensional vector of latent spatial random effects. Then, the spatial GLMM
is specified as
\begin{align}
  & Y(\mathbf{s}_i) \mid \mathbf{X}_i, Z(\mathbf{s}_i) \overset{{\rm ind}}{\sim}
    f(\cdot \mid \mu_i, \boldsymbol{\gamma}), \nonumber \\
  & \mu_i = \E[Y(\mathbf{s}_i) \mid \mathbf{X}_i, Z(\mathbf{s}_i)] =
    g^{-1}(\mathbf{X}^\top_i 
    \boldsymbol{\beta} + Z(\mathbf{s}_i)), \label{eq:glmm_mu}
\end{align}
where~$g$ is a link function, $\boldsymbol{\beta}$ is a~$p$-dimensional regression
coefficient vector, and~$f(\cdot \mid \mu_i, \boldsymbol{\gamma})$ is a probability density or
mass function of a distribution with expectation~$\mu_i$ and an additional
parameter vector~$\boldsymbol{\gamma}$. Thus, conditional on~$Z(\mathbf{s}_i)$
and~$\mathbf{X}_i$, we assume that the responses~$Y(\mathbf{s}_i)$'s are
mutually independent.

The spatial dependence among the
responses~$\mathbf{Y} = {\big[Y(\mathbf{s}_1), \ldots, Y(\mathbf{s}_n)\big]}^\top$ is
introduced by a zero-mean
HGP~$\mathbf{Z} \sim \textrm{HGP} \big\{0, v(\cdot; \boldsymbol{\sigma}), r(\cdot;
\boldsymbol{\delta})\big\}$, where $\boldsymbol{\sigma}$
and~$\boldsymbol{\delta} = \{ \rho, \nu \}$, are the SD and PEC functions' parameters,
respectively. To summarize, the parameters of the spatial GLMM
are~$\boldsymbol{\theta} = {\{\boldsymbol{\beta}^\top, \boldsymbol{\sigma}^\top, \boldsymbol{\delta}^\top,
  \boldsymbol{\gamma}^\top \}}^\top$.

\subsection{Bayesian Inferences}\label{sec:bayest}

Prior distributions for all the model parameters are to be specified to complete
the Bayesian model specification. For fixed effects regression
coefficients~($\boldsymbol{\beta}$), we employ uncorrelated zero-mean Normal priors
with marginal variances~$\mathrm{v}_{\beta}$. In the homoscedastic case,
on~$\sigma$ we place a half-t prior with~$3$ degrees of freedom,
denoted~$t_{+}(3)$~\citep{gelman2006prior}. For the hereroscedastic
model~\eqref{eq:nu}, we set independent Normal priors for~$\alpha_0$
and~$\alpha_1$ with marginal variances~$\mathrm{v}_{\alpha}$. Hereafter, we set the
marginal variances as $\mathrm{v}_\beta = 10^2$ and~$\mathrm{v}_\alpha = 1$ unless
otherwise stated. The PEC dependence is encoded
through~$\boldsymbol{\delta} = \{ \rho, \nu \}$. In geostatistics, smoothness parameters
for correlation functions are typically fixed~\citep{zhang2004inconsistent,
  datta2016hierarchical}. Due to this and the scope of this paper, we specify
separate models for different values of~$\nu$, selected based on the empirical
assessment to ensure positive definiteness of the induced correlation matrices,
and select the best fit for inference. The spatial dependence parameter~$\rho$ is
also inherently difficult to estimate. Therefore, to avoid oversmoothing, we set
an exponential prior in the same spirit of the penalized complexity~(PC)
prior~\citep{simpson2017penalising, fuglstad2019constructing} for
$\rho$. Specifically, we employ an exponential prior for~$\rho$ with
mean~$\lambda_\rho = - \rho_0 / \log(p_\rho)$, where~$\rho_0$ is an upper-bound set to four-fifths
of the largest observed Hausdorff distance, and~$p_\rho$ is the probability
that~$\rho$ exceeds~$\rho_0$. The resulting prior concentrates its mass on lower
values~$\rho$. Note that, the resulting prior is not exactly a PC prior since it is
not placed on the Kullback-Leibler divergence with a simpler model.

The posterior distribution of~$\boldsymbol{\theta}$ and~$\mathbf{Z}$ given the
observed data~$\mathbf{y}$ and~$\mathbf{X}$ is
\begin{equation}
  \label{eq:posterior}
  \pi(\boldsymbol{\theta}, \mathbf{z} \mid \mathbf{y}) \propto
  \pi(\mathbf{y} \mid \mathbf{X}, \mathbf{z}, \boldsymbol{\beta}, \boldsymbol{\gamma})
  \pi(\mathbf{z}  \mid \boldsymbol{\sigma}, \boldsymbol{\delta})
  \pi(\boldsymbol{\theta}),
\end{equation}
where~$\pi(\cdot)$ denotes prior and posterior distributions of its arguments.
Importantly, the distribution placed on the random effects~$\mathbf{z}$ also
functions as a prior, as we draw samples from these latent effects distribution
conditioned on the observed data.

The estimation of the spatial GLMM parameters are obtained from the posterior
distribution defined in Equation~\eqref{eq:posterior}. Since this distribution
is intractable, we draw samples from it using the No-U-Turn Markov chain Monte
Carlo~(MCMC) sampler in \texttt{Stan}~\citep{hoffman2014no,
  stan2023rstan}. Unless otherwise stated, we initialize the parameters with
samples from their respective prior distributions. The latent random effects are
initialized from a standard Gaussian distribution. The number of samples and the
warm-up period of the MCMC algorithm are application dependent. We assess the
convergence of the chains using the split-$\hat{R}$
diagnostic~\citep{vehtari2021rank}. Finally, the parameters' point estimates and
95\% credible intervals~(CI) are obtained as, respectively, the median and
percentiles~($2.5$ and $97.5$, unless otherwise stated) of the marginal MCMC
samples.

We assess goodness-of-fit (GoF) using the
leave-one-out~\citep[LOOIC]{vehtari2017practical} and widely
applicable~\citep[WAIC]{watanabe2010asymptotic} information criteria, where
lower values indicate a better fit. When comparing a single candidate model to
the HGP, we utilize the difference in the GoF metric~$G$ between the HGP and the
given model, denoted by~$\Delta_G$. Consequently, a negative~$\Delta_G$ suggests that the
HGP model is a better fit for the data. The competing methods depend on the type
of application we are working with and will be mentioned in the upcoming
sections.

Predicting spatial phenomena accurately at unobserved locations is crucial in
many applications. The HGP prior for spatial random effects in GLMMs offers a
powerful solution. For a set of~$m$ unobserved
locations~$\mathcal{S}^\ast = \{ \mathbf{s}^\ast_1, \ldots, \mathbf{s}^\ast_m \}$, we define
an~$m$-dimensional variable~$\mathbf{Y}^\ast$ representing the response variable at
these locations. Assuming covariates from Equation~\eqref{eq:glmm_mu} are
available, we can generate predictions accounting for uncertainty using the
posterior predictive distribution of $\mathbf{Y}^\ast$:
\begin{equation}\label{eq:preds}
  \pi(\mathbf{y}^{\ast} \mid \mathbf{y}) = \int \pi(\mathbf{y}^{\ast} \mid
  \mathbf{z}^{\ast}, \boldsymbol{\theta})
  \pi(\mathbf{z}^{\ast} \mid \mathbf{z}, \boldsymbol{\theta})
  \pi(\boldsymbol{\theta}
  \mid \mathbf{y}) \, {\rm d} \mathbf{z}^{\ast} \, {\rm d} \mathbf{z} \,
  {\rm d} \boldsymbol{\theta},
\end{equation}
where~$\mathbf{z}^\ast = {(z(\mathbf{s}^\ast_1), \ldots, z(\mathbf{s}^\ast_m))}^\top$ is a
realization of the spatial random effect at the new locations,
and~$\boldsymbol{\theta}$ represents the model parameters. Details on sampling from
this distribution are provided in Section~S.3 of the Supplementary Material.

\section{Simulation Studies}\label{sec:sim}

\subsection{Areal Data}\label{sec:sim_areal}

We conducted a simulation study to assess the performance of the HGP relative to
the DAGAR model, a state-of-the-art model designed explicitly for areal data
\citep{datta2019spatial}, in modeling data generated from other models. For
details about the DAGAR model, we refer to the Appendix. We grounded our
simulations in a real-world context by utilizing the 134 Intermediate Zones~(IZ)
north of the River Clyde within the Greater Glasgow in Scotland (Figure~S.5),
mirroring the application detailed in the upcoming
Section~\ref{sec:data_areal}. This region exhibits a range of Hausdorff
distances between zones, from~$0.7$ to~$37.7$ kilometers~(median: $9.4$
km). Data were generated under a Poisson GLMM with a log link function,
incorporating a single explanatory variable drawn from a standard normal
distribution. The intercept and regression coefficients associated with this
explanatory variable were fixed at~$4.28$ and~$0.33$, respectively, to mirror
the parameters estimated in the real-world
application~(Section~\ref{sec:data_areal}).

Two random effect structures were considered: the Aggregated Gaussian
Process~(AGP) and the Besag--York--Molli\'{e}~\citep[BYM]{besag1991bayesian}, with
model parameters corresponding to their fits to the real data. Details about
those models and their parameters are given in the Appendix,
Section~\ref{app:specialized}. The AGP assumes a point-level GP aggregated at
the areal unit level and is closer in structure to the HGP model. We chose AGP's
model parameters to reflect the HGP results from our real-world
application~(Section~\ref{sec:data_areal}): a point-level marginal standard
deviation~($\sigma$) of~$0.19$, a smoothness parameter~($\nu$) of~$1$, and a range
parameter~($\rho$) of~$2.21$. The BYM model allows for spatially correlated random
effects with a structure closer to that of the DAGAR model. We set its marginal
standard deviation~\citep{riebler2016intuitive}, denoted~$\sigma$, to~$0.24$ and the
spatial dependence parameter~($\xi$) to~$0.43$, aligning with estimates from the
real data analyses in Section~\ref{sec:data_areal}. In Section~S.5.2 from the
Supplementary Material, we detail how to simulate data from the AGP.

We generated 200 datasets from the Poisson GLMM under each of the two distinct
random effect structures. For each simulated dataset, we fit two Bayesian
Poisson GLMMs, the first one using an homoscedastic HGP while the second one
employing a DAGAR prior for the random effects. Both specified the fixed effects
correctly with the single covariate. The priors for the parameters in the HGP
model were set to be those presented in Section~\ref{sec:bayest}. The priors for
the parameters in the DAGAR model were set to the same as those in
\citep{datta2019spatial}; For more details, see the Supplementary Material,
Section~S.4.1. The GoF was evaluated using the criteria in
Section~\ref{sec:bayest}. In the Supplementary Material, Section~S.4.2, we also
evaluated the HGP's ability to recover true model parameters under correct
specification.

\begin{figure}[tb]
  \centering
  \includegraphics{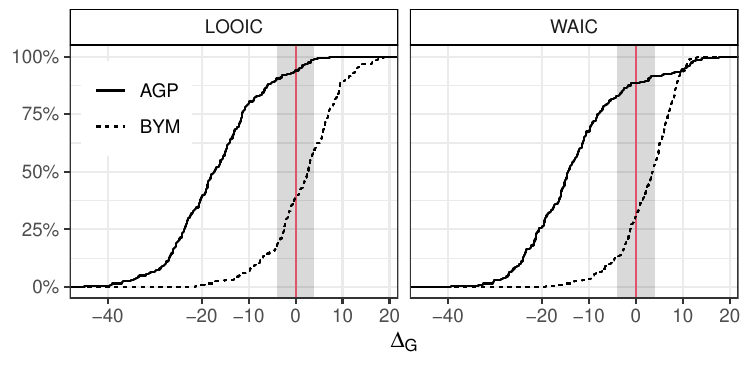}
  \caption{ECDFs displaying the GoF comparison between HGP and DAGAR models
    across different data generating processes (AGP\@: solid line, BYM\@: dashed
    line) and model comparison criteria (columns of the panel) based on 200
    simulated datasets. Shaded region indicates negligible differences. Negative
    $\Delta_G$ indicates better HGP performance.}
  \label{fig:areal_sim_gof}
\end{figure}

Figure~\ref{fig:areal_sim_gof} shows the HGP consistently fitting the data
better than the DAGAR model when data were generated from the AGP\@. The figure
displays empirical cumulative distribution functions (ECDFs) of the differences
($\Delta_G$) in LOOIC and WAIC (left and right columns of the panel, respectively)
between the models. While no formal thresholds exist for significant differences
in the model comparison criteria we used, we consider differences exceeding 5 to
be noteworthy, following a similar practice used with the
DIC~\citep{vranckx2021stability}. For AGP-generated data (solid line), our
simulation study provides strong evidence that the HGP outperforms the DAGAR in
GoF. Specifically, $\Delta_{\textrm{LOOIC}}$ was below $-10$ for 80\% of simulated
datasets, strongly favoring the HGP\@. In contrast, evidence favoring
DAGAR~($\Delta_{\textrm{LOOIC}} > 4$) was observed in only 1\% of the generated
data. Interestingly, the HGP remained competitive when we generated data from
the BYM model~(a structure closer to the DAGAR). While there was some evidence
favoring the DAGAR, $\Delta_{\textrm{LOOIC}}$ was below~4 for more than 50\% of the
simulations under this scenario. These results highlight the HGP's robustness as
a disease mapping model, demonstrating superior performance when the data aligns
with its structure and remains a viable option even when a specialized
competitor better represents the underlying process.

\subsection{Data Fusion}\label{sec:sim_df}

To assess the predictive performance of the HGP against standard data fusion
models, we conducted a simulation study mirroring usual data fusion
scenarios. Within a square region with an area of $51^2$, we generated a map
comprising a $17\times17$ grid-boxes of areal units~(each with an area of $3^2$) and
35 point-referenced data points. Additionally, we created five extra spatial
units for each of the following three types exclusively to evaluate predictions:
larger polygons~(grid boxes with area of $3^2$), smaller polygons~(grid boxes
with area of $1.5^2$), and point-referenced, as displayed in Figure~S.6 from the
Supplementary Material. We carefully chose the study region and grid box
dimensions so each polygon's area exceeded one. This design emulates scenarios
where areal units, often derived from satellite or computer models, are more
abundant than point-referenced data. Moreover, the grid-boxes of different sizes
used to assess the predictions have the purpose of investigating how the
predictions deteriorate by decreasing the size of the spatial units.  Hausdorff
distances between sample units in this map ranged from $0.4$ to $69.8$. We
simulate the datasets from a Normal GLMM with an identity link function, no
covariates, and a spatial random effect. The overall mean of this GLMM was fixed
at $6.28$ to mirror the application from Section~\ref{sec:data_fusion}. In
addition, this model introduces the small scale variance~(denoted $\tau^2$).

This section's spatial GLMMs employ two distinct spatial random effect
structures: a heteroscedastic HGP and an AGP. The HGP's SD
function~$v(\mathbf{s})$ is defined
as~\(\exp \{ \alpha_0 + \alpha_1 \mathbbm{1}(\mathcal{A}(\mathbf{s}) > 0) \} \), allowing for
distinct variances between point-referenced and areal data. While the AGP is a
zero-mean second-order stationary GP with an isotropic Mat\'{e}rn covariance
function with smoothness parameter $\kappa = 1$. See the Appendix for further details
on the AGP model, and \citet{moraga2017geostatistical} for a detailed definition
of the Mat\'{e}rn covariance function.

We selected HGP parameters~($\tau^2 = 0.17^2$, $\sigma = 4.4416$,
$\sigma_a = 1.3114$, $\rho = 28.45$, and $\nu = 0.8$) to mirror what was estimated in the
data analysis presented in Section~\ref{sec:data_fusion}. Note that the spatial
variance for point and areal units~($\sigma$ and $\sigma_a$, respectively) is a one-to-one
transformation of ${(\alpha_0, \alpha_1)}^\top$. We set the dependence
parameter~$\rho$ to~48.9. This value differs from our estimate in the real data due
to a necessary rescaling based on the map used in this simulation study. For
further details, refer to the Supplementary Material, Section~S.5.1. We
simulated AGP data with parameters $\sigma^2 = 2.71^2$, $\rho = 2.29$, and
$\tau^2 = 1\times{10}^{-16}$, again mirroring the real data analysis in
Section~\ref{sec:data_fusion}. In aggregated models, $\tau$ approaches zero as only
the average of homoscedastic independent residuals is observed, which converges
in probability to zero.

We then generated 200 datasets from the Normal GLMM under each of the two
different spatial random effects' structures detailed above. For each simulated
dataset, we fit three Bayesian Normal GLMMs for each simulated dataset, one
using the aforementioned heteroscedastic HGP prior for the random effect and
other two using the AGP prior. When fitting the HGP model, we used the priors
highlighted in Section~\ref{sec:bayest}. In addition, we put a penalized
complexity prior~\citep{simpson2017penalising} on $\tau^{-2}$, such that
$\mathbbm{P}(\tau > a) = .05$, where $a$ is half the empirical SD observed in the
data. Inferences and predictions were based in 2000 MCMC samples with a 1000
warm-up phase per chain with four parallel chains. Inferences for GLMMs with an
AGP model are detailed in Section~S.5.3 from the Supplementary Material.

An important aspect of the AGP model lies in the approximation of the stochastic
integrals' from Equation~\eqref{eq:integral}. They are usually approximated
numerically~\citep{fuentes2005model}. In order to show the inferences and the
predictions may change substantially depending on the accuracy of these
approximations, we considered a two different grids to compute those integrals
numerically: a sparse-~(denoted, AGP\textsubscript{1}) and a
fine-resolution~(denoted, AGP\textsubscript{2}) grid, with approximately~2
and~25 points per polygon, respectively. Their priors are detailed in
Section~S.5.3 of the Supplementary Material. In Supplementary Material,
Section~S.5.4, we also present the results for parameter estimation under the
correct specification of the the HGP model.

We assess the quality of model predictions across different scenarios using
three metrics. Firstly, point and interval predictions are derived from the
posterior predictive distributions~(Equation~\eqref{eq:preds}), using their
median and the 2.5\% and 97.5\% percentiles, respectively. The root mean square
error of prediction~(RMSP) is used to evaluate point estimates. The remaining
two metrics assess interval predictions: the frequentist coverage percentage of
prediction intervals, denoted CPP, and the interval
score~\citep[IS]{gneiting2007strictly}. The latter is defined as
\begin{equation*}
  {\rm IS}(y; l, u) =
  u - l  +
  \frac{2}{\alpha} (l  - y) \mathbbm{1}\{ y < l \}
  + \frac{2}{\alpha} (y - u) \mathbbm{1}\{ y > u \},
\end{equation*}
where~$l$ and~$u$ are the endpoints of the symmetric $(1 - \alpha) \%$
prediction intervals for the target~$y$.

\begin{table}[t]
  \centering
  \caption{Out-of-sample prediction for the three models under different data
    generating models. Fit specifies the model fitted to the data. RMSP is the
    root mean squared error of prediction, CPP is the frequentist coverage
    percentage of the 95\% prediction interval, and IS is the interval
    score. Double headers denote the true data-generating
    model.}\label{tab:pred_df}
\addtolength{\tabcolsep}{-1pt}
\begin{tabular}{crrrrrr}
  \hline
  & \multicolumn{3}{c}{AGP} & \multicolumn{3}{c}{HGP} \\ 
  \cline{2-4} \cline{5-7}
  Fit & RMSP & CPP & IS & RMSP & CPP & IS \\ 
  \hline
  AGP\textsubscript{1} &         1.56  & 67.8 & 17.74         & 1.81 & 57.9 & 31.37 \\ 
  AGP\textsubscript{2} & \textbf{1.39} & 95.7 & \textbf{6.60} & 1.80 & 88.1 & 10.63 \\ 
  HGP                  &         1.65  & 90.8 & 8.65          & \textbf{1.51} & 95.7 & \textbf{7.05} \\
  \hline
\end{tabular}
\end{table}

Table~\ref{tab:pred_df} presents the prediction assessment metrics under both
data-generating models. When the AGP is the true model, the fine mesh AGP model
yields the most accurate point predictions~(lowest RMSP). However, the HGP
model's performance is only marginally worse, with a RMSP 18\% higher. For
interval predictions, the HGP model produces prediction intervals~(PI) with
near-nominal CPP and competitive IS. The AGP\textsubscript{1} model suffers from
poor coverage and, consequently, a high IS\@. As expected, when the true
data-generating model is the HGP, the HGP model's predictive performance
surpasses that of its competitors. The point predictions from both AGP models
are remarkably similar, with the fine-mesh AGP exhibiting a RMSP only 19\%
higher than the correctly specified HGP\@. This suggests the fine-mesh AGP
remains competitive in terms of point predictions, despite model
misspecification. In terms of interval predictions, while the
AGP\textsubscript{2} was competitive, its sparse mesh counterpart delivered poor
interval predictions. The performance of the competing method varies
considerably depending on the mesh resolution, which can significantly impact
prediction accuracy. In contrast, our model eliminates the need for arbitrary
mesh selection and consistently delivers highly competitive predictions,
enhancing prediction accuracy. Additional analyses assessing predictions across
different spatial resolutions are presented in the Supplementary Material,
Section~S.5.4.

\section{Applications}\label{sec:data}

\subsection{Areal Data: Respiratory Disease  Hospitalization}
\label{sec:data_areal}

We first analyze a dataset on the count of hospital admissions due to
respiratory disease in 2010 for the 134 intermediate zones (IZ) to the north of
the river Clyde in the Great Glasgow and Clyde health board in Scotland. This
dataset is publicly available in the \texttt{CARBayesdata}
package~\citep{lee2022carbayesdata}. For each IZ, the expected number of
admissions based on the national age- and sex-standardized rates, alongside the
percentage of people classified as income deprived in each IZ are also
available. The standardized morbidity ratio~(SMR), the ratio between the number
of observed and expected cases, is often used to control for confounding factors
when estimating risks~\citep{greenland1984bias}. The goals of the analyses are:
(1) to provide smooth estimates of the SMR, adjusted for the percentage of
income deprivation, and (2) to contrast the outcomes from the proposed
methodology with other established methods.

In this application, $Y(\mathbf{s}_i)$ is the number of admissions by
respiratory disease registered at the $i$-th IZ, denoted $\mathbf{s}_i$. In the
$i$-th IZ, define $x_i$ as the percentage of people who are defined to be income
deprived and $E_i$ as the expected number of admissions. We analyze the data
using the GLMM model from Section~\ref{sec:spglm}, with a Poisson likelihood, a
$\log$ link function, an intercept, a covariate~$x_i$, and an offset~$E_i$. We
employed the homoscedastic HGP~(with a PEC function with smoothness $\nu = 0.7$)
as the prior for the random effects; the choice of~$\nu = 0.7$ is justified in
Table~S.4 of the Supplementary Material. The priors and initialization scheme
for the parameters in the HGP GLMM match those presented in
Section~\ref{sec:bayest}. Inference was performed using four parallel chains
with 5000 MCMC samples, thinned by~10, after discarding the initial 1000
iterations as the warm-up period. The highest split-$\hat{R}$ was smaller than
$1.01$, providing evidence of convergence.

Two competing models, BYM and DAGAR~(see the Appendix for further details), were
considered as the priors for the spatial random effects~$\mathbf{z}$ for
comparison. In the BYM model, parametrized by the marginal
precision~\citep{riebler2016intuitive}, we followed \citet{morris2019Bayesian}
and set standard normal priors for the intercept~($\beta_0$), regression
coefficient~($\beta_1$), and the $\mathrm{logit}$ of the mixing
parameter~$\xi$. Similarly, a standard normal distribution truncated on positive
real numbers was used for~$\sigma$. The priors for the DAGAR model are the ones
presented in Section~S.4.1. The initialization of the MCMC algorithm followed
the same principle used for the HGP model. In both cases, we used four parallel
chains with a warm-up period of 4000 iterations and based the inferences on
20000 MCMC samples, thinned by~20. For these models, the parameters
split-$\hat{R}$ provided evidence of convergence. In addition, trace and density
plots for posterior samples of all fitted model parameters are provided in
Section~S.6.1 of the Supplementary Material.

\begin{table}[t]
  \centering
  \caption{Parameter estimates and model comparison criteria for the three
    models fitted to the areal dataset. Marginal posterior summary through
    median and 95\% credible intervals.}\label{tab:summary_areal}
  \addtolength{\tabcolsep}{-1pt}
  \begin{tabular}{lccc}
    \hline
    & HGP & DAGAR & BYM \\ 
    \hline
    $\beta_0$ & $-$0.21 ($-$0.268, $-$0.139) & $-$0.26 ($-$0.450, $-$0.137) & $-$0.22 ($-$0.254, $-$0.184)\\
    $\beta_1$ & 0.33 (0.284, 0.368) & 0.31 (0.258, 0.370) & 0.33 (0.285, 0.366)\\
    $\sigma$ & 0.19 (0.155, 0.234) & 0.30 (0.218, 0.484) & 0.24 (0.176, 0.354)\\
    $\rho$ & 2.25 (0.159, 6.948) &  & \\
    $\psi$ &  & 0.43 (0.069, 0.827) & \\
    $\xi$ &  &  & 0.42 (0.100, 0.815)\\
    &  &  & \\
    LOOIC     & \bf{1081.0} & 1081.9 & 1089.3\\
    WAIC      & 1038.0 & \bf{1032.4} & 1039.6\\
    \hline
  \end{tabular}
  \addtolength{\tabcolsep}{1pt}
\end{table}

We present the posterior medians and 95\% CIs for the parameters from the HGP,
DAGAR, and BYM models in Table~\ref{tab:summary_areal}. When considering the
coefficient linked to the covariate representing income deprivation, all three
models yield similar posterior estimates. All models found a positive
association between income deprivation and hospitalizations~($\beta_1$). However,
the HGP provided a more precise and interpretable estimate of spatial dependence
via its practical range parameter, whereas the credible intervals for the
dependence parameters in the BYM and DAGAR models were very wide, suggesting
inconclusive evidence of spatial clustering.

To gain a better understanding of this value in the context of the data, in the
Supplementary Material, Figure~S.9 displays a density histogram of all pairwise
Hausdorff distances observed in the dataset. The plot also features the prior
distribution~(in red) and estimated posterior density~(in blue) associated with
the practical range. Notably, the majority of the posterior distribution mass
for~$\rho$ falls below the concentration of most pairwise distances, indicating a
negligible clustering effect in hospital admissions. An additional analysis of
the spatial correlation is presented in Section~S.6 in the Supplementary
Material.

In Table~\ref{tab:summary_areal}, we also present two model comparison criteria
described in Section~\ref{sec:sim} for the models fitted to the data. These
results suggest the HGP stands out as the best fit based on the LOOIC, rendering
the HGP prior a competitive fit for these data. Despite the modest improvement
in model fitting capacity, the HGP possesses an advantage in terms of intuitive
interpretability of its spatial dependence, achieved in the last two paragraphs
through the practical range parameter $\rho$. Among the specialized models for
areal data, the BYM was the least adequate in this application. However, none of
the models delivered a remarkably better fit to this dataset. This is further
reinforced by looking at the maps of observed and smoothed SMR estimates
presented in Figure~S.8 from the Supplementary Material.

\subsection{Data fusion: Air pollution in Ventura and Los
  Angeles}\label{sec:data_fusion}

Our second application analyzes fused fine particulate
matter~(PM\textsubscript{2.5}) data in Ventura and Los Angeles counties,
California, USA\@. The dataset combines point-referenced direct measurements
\citep{epa2022clean} with areal satellite-derived estimates
\citep{van2015global}, averaged from 2010 to 2012~(Supplementary Material,
Figure~S.23). To align temporally, we averaged the direct measurements over the
same period. The study region comprises 19 measurement stations and 184 grid
boxes~(approximately 101.95 km\textsuperscript{2} each), excluding islands. Our
primary goal is to predict PM\textsubscript{2.5} concentrations at a finer
scale~(approximately 45 km\textsuperscript{2}) using both data
sources. Secondarily, we aim to estimate average PM\textsubscript{2.5}
concentration and its spatial dependence across both counties.

Building on previous concepts, we introduce the notation for this
application. The response variable is the recorded PM\textsubscript{2.5},
denoted $Y(\mathbf{s}_i)$. In this case, $\mathbf{s}_i$ is either a point or a
grid box. We again resort to the GLMM model from Section~\ref{sec:spglm}, but
without any covariates. We assume the likelihood is Normal with variance
$\tau^2$, and the link function is the identity. A heteroscedastic HGP prior is
placed on the random effect allowing for differing variances between
point-referenced and areal data through the SD function from
Equation~\eqref{eq:nu} with
$w(\mathbf{s}) = \mathbbm{1}(\mathcal{A}(\mathbf{s}) > 0)$. For interpretability, we
present results for $\sigma = \exp\{ \alpha_0 \}$ and
$\sigma_a = \exp\{ \alpha_0 + \alpha_1 \}$ instead of $\alpha_0$ and $\alpha_1$. The model was fit with
$\nu$ fixed at different values, and we selected the one with best
goodness-of-fit~($\nu = 0.7$). For more details, see Table~S.5 in the
Supplementary Material. The priors for model parameters are as in
Section~\ref{sec:sim_df}. We used four parallel chains, each comprising a
warm-up period of 1000 samples and an additional 1000 samples for inference. All
model parameters exhibited split-$\hat{R}$ values below~1, indicating good
convergence. This is further supported by the trace plots presented in
Section~S.7.1 of the Supplementary Material.

For comparison, we also fit two models with AGP random effects, with priors for
the model parameters as described in Section~\ref{sec:sim_df}. The two models
only differ in the precision of the numerical integrals approximation. The
sparse- and fine-resolution AGP models are denoted AGP\textsubscript{1} and
AGP\textsubscript{2}, respectively.

\begin{table}[t]
  \centering
  \caption{Parameter estimates and out-of-sample prediction assessment for the
    three models fitted to the fused dataset. Marginal posterior summary through
    median and 95\% credible intervals. RMSP is the root mean squared error of
    prediction; CPP is the frequentist coverage of the 95\% prediction interval;
    and IS is the interval score.}\label{tab:summary_df}
  \begin{tabular}{lccc}
    \hline
     & HGP & ${\rm AGP_1}$ & ${\rm AGP_2}$ \\ 
    \hline
    $\beta$   & 5.593 (4.659, 6.431) & 6.176 (2.160, 9.990) & 6.344 (6.102, 6.585)\\
    $\rho$   & 15.076 (8.445, 28.561) & 13.093 (4.809, 29.720) & 0.493 (0.340, 0.750)\\
    $\tau$   & 0.174 (0.078; 0.324) & 1.403 (1.253; 1.573) & 0.539 (0.394, 0.719)\\
    $\sigma$   & 3.893 (2.962; 5.359) & 1.717 (0.933; 2.938) & 2.877 (2.144; 3.566)\\
    $\sigma_a$ & 1.275 (1.064; 1.590) & & \\
    & & & \\
    RMSP      & \textbf{2.22} & 3.55 & 4.45 \\
    CPP       & 95.6          & 77.3 & 93.6 \\
    IS        & \textbf{5.08} & 15.4 & 9.55 \\
    \hline
  \end{tabular}
\end{table}

We present the posterior median and 95\% CI of the model parameters in
Table~\ref{tab:summary_df}. When looking at the common parameters between the
three models~(namely, $\beta_0$, $\sigma$, $\rho$, and $\tau$), the HGP acts as a compromise
between AGP\textsubscript{1} and AGP\textsubscript{2} while eliminating the
arbitrary choice of the level of discretization for approximating numerical
integrals. Unlike the aggregated models, our method allows the data to tell us
whether the variance associated to areal units is smaller or larger than that of
point-referenced data. Interestingly, the estimated SD of the point-referenced
data is 3.09 times larger than the areal data (95\% CI:\@ 2.43, 4.12). The
interpretation of $\rho$, presented in tens of kilometers, in both models is
carried out in a similar fashion because the Hausdorff distance reduces to the
Euclidean distance between points at the point-level.

Additionally, Table~\ref{tab:summary_df} presents the out-of-sample prediction
performance of the three models, assessed via 10-fold cross-validation. Using
the three metrics from our prior simulation study~(RMSP, CPP, IS), we find the
HGP offers the most reliable point predictions, evidenced by its lowest
RMSP\@. The HGP also maintains nominal coverage for its prediction
intervals. While the AGP\textsubscript{2} achieves similar coverage, its
interval scores are considerably higher. Table~S.6 in the Supplementary Material
provides a breakdown by data type, highlighting the poor performance of
competing methods for point-referenced data. The PI frequentist coverage, for
example, can be as low as 3\%.

In Table~S.7 of the Supplementary Material, we further compared the predictive
performance of those models to two additional alternatives: a Log-Normal model
with HGP random effects, and a GP where the areal units are reduced to their
centroids, resulting in a geostatistical model. We have not tested an AGP where
the response follows a Log-Normal distribution, because the random effects would
need to be aggregated on the response scale~\citep{suen2025coherent}, which can
be seen as a consequence of Jensen's inequality. Surprisingly, the AGP models
performed the worst, while the Log-Normal HGP provided the second-best
predictive performance.

The results from the AGP models diverged significantly depending on the mesh
resolution. The credible interval width for the intercept varies significantly
between the two models. The AGP\textsubscript{1} model indicates moderate
spatial dependence, while AGP\textsubscript{2} suggests weak dependence. This
weak dependence may imply AGP\textsubscript{2} assigns nearly independent random
intercepts to each data point, with variances inversely proportional to the
sample unit's area. AGP mesh resolution is typically practitioner-informed, and
results are highly sensitive to this tuning parameter, which lacks a consensus
on ``optimal resolution'' \citep{fuentes2005model, liu2011empirical,
  wang2021combining}.

\begin{figure}[t]
  \centering
    \includegraphics{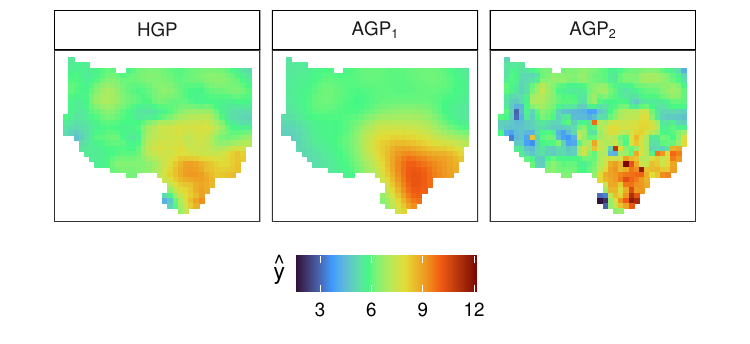}
    \caption{Change-of-support of PM\textsubscript{2.5} to a finer resolution
      using the sparse and fine AGP and the HGP priors.}\label{fig:pred_maps_df}
\end{figure}

Figure~\ref{fig:pred_maps_df} displays the change-of-support predictions
provided by the three models we fitted to the data. HGP predictions indicate
higher PM\textsubscript{2.5} concentrations in the southeast~(Los Angeles) and
provide precise estimates with uncertainty quantification in areas lacking
measurement stations, such as the northwestern part of the map. Unlike its
counterparts, the HGP's ability to generate parsimonious predictions without
arbitrary discretization, combined with its interpretable parameters, makes it
an attractive alternative for spatial data fusion problems.

\section{Discussion}\label{sec:disc}

The proposed framework unifies the modeling of spatial data, extending
geostatistical models to include areal and fused data within the same
setting. Despite the mathematical simplicity, the subtle change of paradigm
powered by the HGP opens a new direction of research in spatial statistics while
taking advantage of what has been proposed in the geostatistics and GP
literature. The proposal has the potential to simplify how we deal with complex
problems such as spatial misalignment and change of support. In simpler
scenarios, as a prior for spatial random effects in a GLMM, the HGP proves to be
competitive to models specially designed for areal and fused data types. Our
simulation studies indicate that the HGP is particularly advantageous for areal
data with units of varying sizes and shapes, as confirmed by our disease mapping
application. Unlike the traditional areal models, the HGP also provides a
directly interpretable spatial parameter, offering insights into the practical
range of spatial dependence. In data fusion contexts, the HGP delivers more
reliable interval predictions compared to the sparse AGP, and it provided
similar results to the fine AGP in parameter estimation and spatial
range. Additionally, in the air pollution data application, the proposed model
consistently outperformed the AGP regardless of the resolution of its mesh. This
success is even more remarkable considering the HGP bypasses the subjective step
of defining a grid or mesh, a crucial but problematic requirement for the AGP.\@

The empirical assessment in Section~\ref{sec:corr} reveals an interesting
insight: the properties of the induced process appear linked to the geometric
complexity of the sets involved. This is most evident in the contrasting results
from our two scenarios. For the data fusion scenario, which involves only convex
sets, all correlation matrices were positive definite. In contrast, for the
areal data scenario, where most spatial units are non-convex, we observed
negative eigenvalues at the border of the parameter
space~(Figure~\ref{fig:corr-exp}). This contrast provides strong empirical
evidence that the PEC function is a valid and robust choice for applications
involving convex sets, such as most data fusion problems.

While our extensive empirical assessments consistently demonstrate the positive
definiteness of the PEC function across practical parameter ranges, a formal
proof establishing the general conditions under which functions of the Hausdorff
distance are positive definite remains an open and challenging theoretical
question. Nevertheless, the proposed HGP addresses several long-standing
problems in spatial statistics. For instance, traditional areal models are
unsuitable for out-of-sample prediction and often have counter-intuitive
parameters interpretation~\citep{wall2004close,
  assuncao2009neighborhood}. Furthermore, the validity of CAR models, for
instance, also depends on constraining the parametric space to ensure a
non-singular precision matrix~\citep{wall2004close}. While data fusion models do
not share the same issues, we have shown that their results can be significantly
affected by the arbitrary precision of required approximations. By introducing
the HGP and demonstrating its utility, we aim to provide a foundation for future
theoretical work, thereby bridging the gap between practical application and
complete mathematical understanding.

Several conceptual challenges for the HGP remain open for future
investigation. Relaxing the isotropy assumption, for instance, requires further
theoretical development. To apply the HGP to data on manifolds like the sphere,
new results are needed to guarantee the process's validity. Furthermore, because
the HGP is a set-valued function, any analysis of its
curvature~\citep[e.g.,][]{halder2024Bayesian} necessitates more general concepts
of differentiation~\citep{khastan2021new}. It is important to note, however,
that this latter complexity does not prevent the use of powerful inference
frameworks that rely on the posterior's gradient, such as variational
inference~\citep{blei2017variational}.

Although the~$\mathcal{O}(n^3)$ computational complexity is a limitation, the method was
developed with the intent of leveraging what has already been developed for
GPs. Thus, adapting scalable nearest neighbor
approximations~\citep{datta2016hierarchical, quiroz2023fast} for the HGP is a
viable avenue for future research. The main challenge would be to identify
nearest neighbors in the Hausdorff distance sense while avoiding the storage of
a $n \times n$ distance matrix.

\subsubsection*{Conflict of interest}

All authors declare that they have no conflicts of interest.

\bibliographystyle{chicago}
\bibliography{references}

\appendix

\section{Review of Existing Methods}\label{app:specialized}

The BYM model we use corresponds to a zero-mean multivariate Normal model with
covariance matrix defined as
\[
  \sigma^{-2}((1 - \xi) \bm{I} + \xi \bm{Q}^{-}),
\]
where~$\sigma$ is the average marginal SD, $\bm{I}$ is the identity matrix, and
$\bm{Q}^{-}$ is the generalized inverse of the scaled precision matrix of an
ICAR model. The spatial dependence parameter $\xi$ is a mixing parameter. Its
parametric space comprises of real numbers between 0 and 1, with larger values
indicate stronger spatial dependence. The entries of~$\bm{Q}$ are as follows:
\[
  Q_{ij} =
  \begin{cases} k n_{N(i)}, & \text{ if } i = j, \\
    -k, & \text{ if } i \sim j, \\
    0, & \text{ elsewhere,}
  \end{cases}
\]
where~$n_{N(i)}$ denotes the number of neighbors for the region~$i$,
and~$i \sim j$ indicates that $j$ and $i$ are neighbors, and~$k$ is the scaling
constant. The scaling constant depends on the adjacency matrix and enhances the
interpretation of~$\sigma$, which becomes the average marginal standard deviation, as
opposed to the conditional standard deviation in the original
formulation~\citep{riebler2016intuitive}. We used the BYM as a prior for the
random effects in the spatial GLMMs~(Section~\ref{sec:spglm}) for areal data in
Section~\ref{sec:data_areal}. We also generated data from a spatial GLMM
assuming the BYM was the actual distribution of the spatial random effects for
part of the simulation studies in Section~\ref{sec:sim_areal}.

The DAGAR model, in our context, is a zero-mean multivariate Normal model. We
used the DAGAR as a prior for the spatial GLMM~(Section~\ref{sec:spglm}) random
effects~(e.g., $\mathbf{z}$) for areal data, in particular, in
Section~\ref{sec:sim_areal} and~\ref{sec:data_areal}. The precision matrix for
the DAGAR model is constructed using the data ordering, an adjacency matrix, and
two parameters: a conditional precision, $\tau$, and a spatial dependence
parameter, $\psi$.  The~$\psi$ parameter represents the average spatial correlation
between adjacent spatial units. Specifically, the data ordering and adjacency
matrix define a tree graph, and the precision matrix is derived from an order-1
autoregressive process on the nodes of this graph. A key advantage of this model
is its computational efficiency, as it can be evaluated using either closed-form
conditional distributions or direct formulas for the Cholesky decomposition of
its covariance matrix. In particular, the Cholesky decomposition of the
precision matrix under a DAGAR model can be written as
$\mathbf{F}^{-1 / 2}(\mathbf{I} - \mathbf{B})$, where $\mathbf{F}$ is a diagonal
matrix, $\mathbf{I}$ is the identify matrix, and $\mathbf{B}$ is
lower-triangular~\citep{datta2022nearest}. The entries of $\mathbf{F}$ and
$\mathbf{B}$, respectively, are given by
\begin{align*}
  F_{ii} & = {[\tau (1 + 1 + n_{\vec{N}(i)} \psi^2)]}^{-1} \quad \text{ and } \\
  B_{ij} & = \mathbbm{1}(j \sim i, j < i) \frac{\psi}{1 + 1 + n_{\vec{N}(i)} \psi^2},
\end{align*}
where $\mathbbm{1}(A)$ denotes an indicator function, equaling~$1$ if the
event~$A$ happens and~$0$ otherwise. Note that, in this
case,~$n_{\vec{N}(i)} = \sum_j \mathbbm{1}(j \sim i, j < i)$ represents the number of
``directional neighbors''. Although the DAGAR depends on an arbitrary order for
the data points, the authors also developed an ``order-free'' variant of the
model. However, the ``order-free'' DAGAR does not enjoy the computational
efficiency of its ordered counterpart.

The AGP model is employed as a prior on the random effects~($\mathbf{z}$) of the
spatial GLMM~(Section~\ref{sec:spglm}) for fused data. In particular, we used
this model to generate data for the simulation studies in Section~\ref{sec:sim},
and to analyze the dataset from Section~\ref{sec:data_fusion}. The AGP core
assumption is as follows:
\begin{equation} \label{eq:integral}
  Z(\mathbf{s})
  = \begin{cases}
    {\mathcal{A}(\mathbf{s})}^{-1}
    \int_{\mathcal{A}(\mathbf{s})} Z(\mathbf{s}) \dd \mathbf{s}, & \text{if }
                                                      \mathcal{A}(\mathbf{s}) > 0, \\
    Z(\mathbf{s}), & \text{otherwise,}
  \end{cases}
\end{equation}
where~$\{ Z(\mathbf{s}) \, : \, \mathbf{s} \in D \}$ is a~(point-level) zero-mean,
second order stationary and isotropic GP~\citep[p.~53]{cressie1993statistics}
defined over the study region, $\mathcal{A}(\cdot)$~is a function that returns the area of a
spatial object.

\end{document}